\documentclass{pas}

\usepackage{multirow}
\usepackage{amsmath}

\begin{document}

\lefttitle{Publications of the Astronomical Society of Australia}
\righttitle{B. Amend \textit{et al.}}

\jnlPage{1}{xx}
\jnlDoiYr{2026}
\doival{10.1017/pasa.xxxx.xx}

\articletitt{Research Paper}

\title{Outflows in steep density gradients: diversity of behavior and implications for tidal disruption events and luminous fast blue optical transients}

\author{
\gn{Benjamin} \sn{Amend}$^{1}$,
\gn{Michael} \sn{Camilo}$^{2}$,
\gn{Eric R.} \sn{Coughlin}$^{1}$,
\gn{Anna Y. Q.} \sn{Ho}$^{2}$, and
\gn{Jonathan} \sn{Zrake}$^{3}$
}

\affil{
$^1$Department of Physics, Syracuse University, Syracuse, NY 13210, USA\\
$^2$Department of Physics, Cornell University, Ithaca, NY 14850, USA\\
$^3$Department of Physics and Astronomy, Clemson University, Clemson, SC 29634, USA
}

\corresp{Eric R. Coughlin, Email: ecoughli@syr.edu}

\citeauth{Amend B., Camilo M., Coughlin E. R., Ho A. Y. Q., and Zrake J. Wind bubbles in steep density gradients. {\it Publications of the Astronomical Society of Australia} {\bf 00}, 1--xx. https://doi.org/xx.xxxx/xxxxx}

\history{(Received xx xx xxxx; revised xx xx xxxx; accepted xx xx xxxx)}

\begin{abstract}
Powerful explosions may undergo sustained energy injection as a central engine launches a wind into the surrounding gas, generating a forward and a reverse shock separated by a contact discontinuity. During the adiabatic phase, the dynamics depend strongly on the wind-to-ambient density ratio $f \equiv \rho_{\rm w} / \rho_{\rm a}$. For $f \ll 1$, the reverse shock lies well inside the contact discontinuity, and the mechanical energy deposited by the wind is retained in a radially extended, approximately isobaric shocked-wind region whose pressure drives the swept-up ambient shell. For $f \gg 1$, the reverse shock remains close to the contact, and the expansion is governed more directly by the ram-pressure interaction between the freely expanding wind and the swept-up ambient gas. We use analytic scalings and one-dimensional shock-capturing hydrodynamic simulations to determine how outflows in these two limits evolve in ambient density profiles $\rho_{\rm a} \propto r^{-n}$, where $2 \leq n \leq 3$, and whether their shock structures accelerate or coast at constant velocity. For $n > 2$, initially underdense outflows produce accelerating forward shocks whose radii evolve as $R_{\rm s} \propto t^{3/(5-n)}$. Because $\rho_{\rm w} \propto r^{-2}$, $f$ increases with radius, causing the reverse-shocked wind region to contract relative to the contact position as the forward shock transitions toward constant-velocity expansion. This transition occurs when $f \sim$ a few, at a time $t_{\rm dec} \propto f_0^{1/(2-n)}$, where $f_0$ is the initial wind-to-ambient density ratio. By contrast, outflows initialized with $f_0 \gg 1$ do not develop an extended accelerating phase and instead remain approximately coasting throughout their adiabatic evolution. We discuss applications to tidal disruption event outflows and luminous fast blue optical transients, whose environments are often inferred to have steep density profiles with $n > 2$.
\end{abstract}


\maketitle

\section{Introduction}

Wind-driven bubbles form when sustained outflows expand into their surroundings, converting the mechanical power of the wind into thermal energy and radiation. This basic structure underlies a wide range of feedback and transient phenomena, from stellar and cluster winds \citep{1972SvA....15..708A, 1975ApJ...200L.107C, 1975ApJ...195..157C, 1977ApJ...218..377W, 1985Natur.317...44C, 2001A&A...369..574V, 2007ARA&A..45..177C} to galactic outflows \citep{1995ApJ...451..498M, 2005ARA&A..43..769V, 2012MNRAS.425..605F, 2015ARA&A..53..115K, 2016MNRAS.455.1830T} and those from tidal disruption events \citep[e.g.][]{2011MNRAS.415..168S}.

A wind impacting an ambient medium generates a two-shock structure: a forward shock expands into the ambient gas, a reverse shock decelerates the freely expanding wind, and a contact discontinuity separates the shocked ambient material from the shocked wind. In the classic wind-bubble picture, the reverse shock eventually recedes well inside the contact discontinuity once the wind density falls substantially below that of the ambient medium, leaving behind a radially extended region of hot, approximately isobaric material (the shocked wind) that drives the swept-up ambient shell through $p\,dV$ work. This pressure-driven stage was described by \citet{1977ApJ...218..377W} and applies prior to the onset of substantial radiative losses for a constant-density ambient medium, and was generalized by \citet{1992ApJ...388..103K} to ambient density profiles $\rho \propto r^{-n}$. In this regime, basic considerations and assumptions show that the forward-shock radius evolves as $R_s \propto t^{3/(5-n)}$, implying deceleration for $n < 2$, constant-velocity expansion for $n=2$, and acceleration for $n > 2$.

These pressure-driven scalings apply only after the reverse shock has receded to radii well inside the contact discontinuity and the speed of the reverse shock is well below the sound speed. At earlier times, or for sufficiently overdense winds where the distance between the reverse shock and contact discontinuity remains small (relative to either radius) for many wind dynamical times, the shocked wind is not well approximated as an isobaric reservoir. In this ejecta-driven phase, the expansion is governed more directly by the ram-pressure interaction between the freely expanding wind and the swept-up ambient gas, and the forward shock is instead described by the similarity solutions of \citet{2024ApJ...975L..14C}; these solutions were demonstrated in \cite{Amend_Lagomarsino_Coughlin_Zrake_2026} to be upheld for $n<2$. The duration of this phase is brief for winds in uniform-density ambient media, however it can last much longer when the ambient medium has a steep density profile. In particular, the $n=2$ case corresponds to a persistent ejecta-driven state because the ratio of wind to ambient density is constant over time. For $n > 2$, the ambient density decreases faster in radius than the wind density, so the density contrast shrinks over time. As such, if the wind is initialized in the ejecta-driven phase, that phase persists indefinitely, while if the wind is initialized in the pressure-driven phase, it will transition to the ejecta-driven phase at some characteristic time. In the ejecta-driven phase, instead of accelerating, the forward shock (FS) coasts at a velocity marginally greater than the wind speed (which is itself nearly comparable to the CD speed), with the relative separation between the reverse shock (RS) and the contact discontinuity (CD) decreasing with time (see the discussion in \citealt{2024ApJ...975L..14C}). Thus, $n > 2$ does not guarantee acceleration for ejecta-driven bubbles, and the evolution of such systems is dictated by the initial contrast between the wind and ambient densities.

The $n > 2$ regime could apply to a variety of astrophysical phenomena, including galactic superbubbles, tidal disruption events (TDEs), and luminous fast blue optical transients (LFBOTs). Superbubble breakout from stratified galactic disks is associated with a steep ambient density gradient, and the dense swept-up shell can become Rayleigh--Taylor unstable due to its acceleration, fragmenting and allowing hot gas to vent into the halo \citep{1988ApJ...324..776M, 1992ApJ...388..103K, 2017ApJ...834...25K, 2018MNRAS.481.3325F}. When stars are tidally disrupted \citep{1988Natur.333..523R, 2021ARA&A..59...21G}, the material that accretes onto the disrupting black hole can power an outflow that produces an observable electromagnetic signature, and ambient medium density profiles as steep as $\rho \sim r^{-3}$ have been inferred for some of these events \citep{2016ApJ...819L..25A, 2021ApJ...919..127C}. Similarly, the environments of some LFBOTs can be modeled with steep ambient density profiles where $n \geq 2.5$ \citep{2022ApJ...932..116H, 2022ApJ...926..112B, 2025ApJ...993L...6N,2026MNRAS.549ag678P}. For such systems, the radio emission is commonly attributed to synchrotron radiation from shock-accelerated electrons, with synchrotron self-absorption setting the low-frequency turnover \citep{1998ApJ...499..810C}. Because the post-shock magnetic field and characteristic electron energies depend on the shock velocity, accelerating and coasting shocks should produce different evolution of the spectral peak and radio flux, potentially yielding a change in light-curve slope as the system transitions between regimes. These systems therefore motivate an examination of the $n>2$ branch, as the onset, duration, and eventual termination of the accelerating phase may be central to their dynamics and observable signatures.

Here we study the evolution of wind-driven explosions in ambient density profiles with $2 \leq n \leq 3$, where the generalized pressure-driven solutions predict constant-velocity expansion at $n=2$ and acceleration for $n > 2$, but where ejecta-driven bubbles may instead remain in the coasting (i.e., near-constant-velocity) regime. We consider both winds that are initially coasting (with large initial overdensities) and initially accelerating (with the wind initially under-dense relative to the ambient medium). By comparing one-dimensional shock-tracking simulations to the ejecta-driven similarity solutions of \citet{2024ApJ...975L..14C} for $2 \leq n < 3$, and separately examining the limiting $n=3$ case, we delineate regimes in which bubbles in steep density gradients either remain ejecta-driven or transition toward the pressure-driven accelerating behavior described by \citet{1992ApJ...388..103K}.

This paper is structured as follows. In Section~\ref{sec:basic-considerations}, we present the analytic scalings relevant to wind-driven bubbles in steep density gradients, emphasizing the distinction between coasting and accelerating evolution. In Section~\ref{sec:hydrodynamics}, we describe our numerical methods and present our primary simulation results, comparing the resulting shock evolution and fluid profiles to those of the self-similar solutions of \citet{2024ApJ...975L..14C} where applicable. In Section~\ref{sec:discussion}, we interpret these results in the context of bubble acceleration, the transition between hydrodynamic regimes, and possible applications to systems such as superbubble breakout and other outflows encountering steep ambient density gradients. We summarize our main conclusions in Section~\ref{sec:conclusions}.

\section{Basic considerations}
\label{sec:basic-considerations}

We consider a time-steady, spherically symmetric wind with constant velocity $v_{\mathrm{w}}$ and mechanical luminosity $L_{\mathrm{w}} = \dot{M}v_{\mathrm{w}}^2/2$, expanding into a cold ambient medium with a power-law density profile. The density of the freely expanding wind and of the ambient medium are
\begin{equation}
    \rho_{\mathrm{w}}(r) = \rho_{{\mathrm{w}},0}\left( \frac{r}{R_{{\mathrm{w}},0}} \right)^{-2}\,,\hspace{1cm} \rho_{\mathrm{a}}(r) = \rho_{{\mathrm{a}},0}\left( \frac{r}{R_{{\mathrm{w}},0}} \right)^{-n}\,,
\end{equation}
where $R_{{\mathrm{w}},0}$ is a reference radius. The interaction between the wind and the ambient medium produces a two-shock structure consisting of a forward shock at $R_{\mathrm{s}}$ that sweeps up the ambient gas, a contact discontinuity at $R_{\mathrm{c}}$, and a reverse shock at $R_{\mathrm{r}}$ that decelerates the wind.

A useful quantity for characterizing the evolution of the three discontinuities (and the shocked fluids) is the instantaneous ratio of the freely expanding wind density to the ambient density near the contact discontinuity,
\begin{equation}
    f(R_{\mathrm{c}}) \equiv \frac{\rho_{\mathrm{w}}(R_{\mathrm{c}})}{\rho_{\mathrm{a}}(R_{\mathrm{c}})} = f_0\left( \frac{R_{\mathrm{c}}}{R_{{\mathrm{w}},0}} \right)^{n-2}\,,
\end{equation}
where $f_0 \equiv \rho_{{\mathrm{w}},0}/\rho_{{\mathrm{a}},0}$. This density contrast dictates whether the system evolves as a direct interaction between an overdense wind and the swept-up ambient gas, or as a pressure-driven bubble. Specifically, when $f(R_{\mathrm{c}}) \gg 1$, the freely expanding wind is much denser than the ambient medium at the interaction radius. In this limit, the reverse shock remains close to the contact discontinuity, the reverse-shocked wind occupies only a narrow shell, and the expansion is governed primarily by the ram pressure of the wind. This is the ejecta-driven regime described by the self-similar solutions of \cite{2024ApJ...975L..14C}. The relative width of the reverse-shocked wind shell is controlled by the smallness parameter
\begin{equation}
    \delta(R_{\mathrm{c}}) \equiv [f(R_{\mathrm{c}})]^{-1/2}
    = f_0^{-1/2}\left( \frac{R_{\mathrm{c}}}{R_{\mathrm{w},0}} \right)^{(2-n)/2}\,,
\end{equation}
where the $-1/2$ comes from the requirement of pressure continuity across the contact. This ejecta-driven approximation is accurate when $\delta \ll 1$, or equivalently when $f(R_{\mathrm{c}}) \gg 1$.

For $n<2$, the density contrast decreases as the shocked structure expands. An initially overdense wind therefore evolves from an early ejecta-driven state toward a pressure-driven decelerating bubble once $f(R_{\mathrm{c}})$ declines to order unity \cite{Amend_Lagomarsino_Coughlin_Zrake_2026}. Equivalently, $\delta$ increases with radius, so the reverse-shocked wind shell gradually broadens relative to the forward-shocked shell, and the flow can no longer be described as a direct ram-pressure interaction between an overdense wind and the swept-up ambient gas. The duration of the ejecta-driven phase increases with $n$ as
\begin{equation}
    t_{\mathrm{ss,end}} = \frac{R_{w,0}}{v_w}\left( \frac{\Delta_{\mathrm{s}}}{\kappa_{\mathrm{r}}} \right)^{2/(2-n)} f_0^{1/(2-n)}\,, \hspace{0.5cm} n<2\,,
\end{equation}
where $\Delta_{\mathrm{s}}$ and $\kappa_{\mathrm{r}}$ are constants associated with the self-similar solutions described in \cite{2024ApJ...975L..14C}. For $n=2$, this expression formally diverges because the wind and ambient densities have the same radial dependence. In that case, $f(R_{\mathrm{c}})=f_0$ remains constant, so an initially ejecta-driven system with $f_0 \gg 1$ remains ejecta-driven indefinitely.

For $n > 2$, since the ambient density decreases more rapidly than the wind density, $f(R_{\mathrm{c}})$ increases with radius and $\delta$ decreases. Therefore, if the system begins in the ejecta-driven regime, expansion into a steeper-than-wind ambient density profile makes the interaction approximation increasingly robust rather than progressively weaker. Such a system cannot naturally relax into the pressure-driven regime; instead, it remains ejecta-driven as long as the assumed density profiles apply.

In the pressure-driven regime, $f(R_{\mathrm{c}}) \ll 1$, the reverse shock has receded well inside the contact discontinuity, the shocked wind forms a radially extended, hot, approximately isobaric interior, and the swept-up ambient shell is driven primarily by the pressure of this shocked wind reservoir. The pressure-driven scaling \citep{1977ApJ...218..377W, 1992ApJ...388..103K} for the radius of the forward shock is
\begin{equation}
    R_{\mathrm{s}}(t) = \left( \frac{L_{\mathrm{w}}}{\rho_{\mathrm{a},0}R^n_{\mathrm{w},0}} \right)^{1/(5-n)} t^{3/(5-n)}\,,
\end{equation}
giving deceleration for $n<2$, constant-velocity expansion for $n=2$, and acceleration for $n>2$.

For $n>2$, this behavior can only persist while the system remains in the limit $f(R_{\mathrm{c}})\ll 1$. Since $f(R_{\mathrm{c}})$ increases with radius in this accelerating regime, a wind that starts out pressure-driven will eventually evolve out of that regime once $f(R_{\mathrm{c}})$ grows to order-unity values, ceasing to accelerate at some time $t_{\mathrm{dec}}$ and -- presumably, but as we test below -- reaching a constant velocity that is approximately given by the expression in \citet{2024ApJ...975L..14C}.

This deceleration time $t_{\mathrm{dec}}$ can be estimated by setting $f(R_{\mathrm{c}})=1$ and using the pressure-driven scaling of the forward-shock radius, since $R_{\mathrm{s}}/R_{\mathrm{c}} = \mathrm{const}$. This yields
\begin{equation}
    t_{\mathrm{dec}} \sim t_{\mathrm{dyn}}f_0^{1/(2-n)}\,,
    \label{eq:deceleration-time}
\end{equation}
where the dynamical time $t_{\mathrm{dyn}}\equiv R_{\mathrm{w},0}/v_{\mathrm{w}}$.

In the following section, we use hydrodynamic simulations to test both the coasting and accelerating solutions, as well as the transition between them where it occurs, following the evolution of the shock structure in detail.

\section{Hydrodynamic simulations}
\label{sec:hydrodynamics}

\subsection{Numerical Methods and Simulation Setup}

To test the predicted behavior discussed in the previous section, we perform one-dimensional hydrodynamic simulations using the relativistic shock-capturing moving-mesh code described in \cite{Amend_Lagomarsino_Coughlin_Zrake_2026}. This method is particularly well suited to the scenarios examined in this work because the computational domain is restricted to the two shocked regions of the flow: the reverse-shocked wind between $R_{\mathrm{r}}$ and $R_{\rm c}$, and the forward-shocked ambient gas between $R_{\rm c}$ and $R_{\rm s}$. The inner and outer boundaries of the global domain then track the reverse and forward shocks respectively, while the interface between the two subdomains follows the contact discontinuity. The unshocked wind and ambient medium are prescribed analytically at the moving boundaries.

All simulations use a cold, spherical, constant-velocity wind with $v_{\rm w} = 0.01c$, where $c=1$ and relativistic corrections emerge at the order of $v_{\rm w}^2 \sim 10^{-4}$ for this choice of velocity. The wind is launched from $R_{\rm w,0}=1$, and expands into a cold ambient medium with $\rho_{\rm a} = \rho_0 (r/R_{\rm w,0})^{-n}$, where $\rho_0=1$. We take $\gamma=5/3$, and we initialize the shock structure by solving the relevant two-shock Riemann problem \citep{2005MNRAS.363...93Z,Amend_Lagomarsino_Coughlin_Zrake_2026}. Each shocked subdomain is resolved with 800 zones, which is sufficient for numerical convergence (see Appendix B. in \cite{Amend_Lagomarsino_Coughlin_Zrake_2026}). All simulations evolve up to $t=10^{18}t_{\rm dyn}$, where $t_{\rm dyn}= R_{\rm w, 0}/v_{\rm w} = 100$ in code units.

We run two sets of simulations with $n\in[2, 2.25, 2.5, 2.75, 3]$. In the first set, $f_0 = 10^{-3}$, so the wind is initially underdense and the system begins in the pressure-driven limit. In the second set, $f_0 = 10^3$, so the system is initialized in the ejecta-driven limit.

\subsection{Results}

\begin{figure}
\begin{center}
\includegraphics{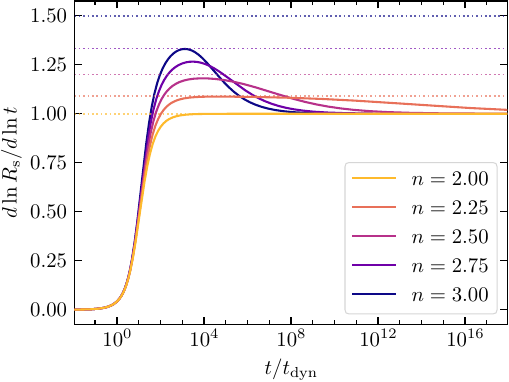}
\caption{Expansion index $\alpha=d\ln{R_{\rm s}}/d\ln{t}$ vs. time for $n \in [2.00, 2.25, 2.50, 2.75, 3.00]$ and $f_0=10^{-3}$. An index $\alpha=1$ indicates constant-velocity expansion, while values above this show acceleration in the forward-shock. The $n=2.5, 2.75, 3$ simulations transition to ejecta-driven behavior before fully relaxing into the pressure-driven regime, as indicated by the gaps between the solid curves and their color-matched horizontal dotted lines.}
\label{fig:expansion-index}
\end{center}
\end{figure}

We first examine the forward-shock dynamics of the initially pressure-driven simulations. Figure~\ref{fig:expansion-index} shows the instantaneous expansion index,
\begin{equation}
    \alpha = \frac{d\ln{R_{\rm s}}}{d\ln{t}}\,,
\end{equation}
for $f_0=10^{-3}$. After an initial 
transient that is related to the initial conditions, the $n>2$ simulations enter an accelerating phase with $\alpha > 1$, consistent with the generalized pressure-driven scaling $R_{\rm s} \propto t^{3/(5-n)}$; the predicted expansion index $3/(5-n)$ is shown by the set of dotted lines in Figure \ref{fig:expansion-index}. As the density contrast increases with radius, 
$\alpha$ gradually relaxes back toward $\alpha \simeq 1$, indicating a transition toward coasting behavior in the ejecta-driven regime.

\begin{figure}
\begin{center}
\includegraphics{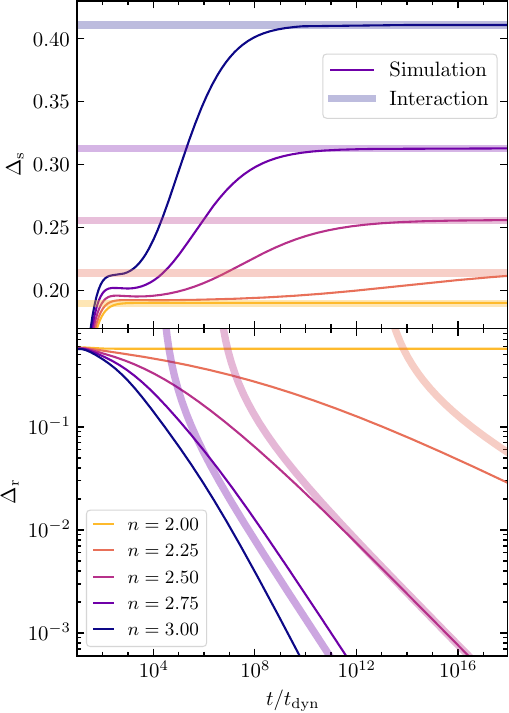}
\caption{The relative thicknesses of the forward- and reverse-shocked shells ($\Delta_{\rm s}$ and $\Delta_{\rm r}$, respectively) vs. time for $f_0 = 10^{-3}$. These simulations all quickly relax into the accelerating limit, then later transition into the coasting limit at $t \sim t_{\rm dec}$. The late time-evolution of the reverse-shocked shell thickness is well-modeled by the solutions in \cite{2024ApJ...975L..14C}, where in the high-$f$ regime, $\Delta_{\rm s} = $ const. and $\Delta_{\rm r} \sim [f(R_{\rm c})]^{-1/2}$. These solutions diverge for $n=3$, so a full time series of $\Delta_{\rm r}$ is not included here. Additionally, convergence slows as $n \rightarrow 3$; this can also be seen in Figure~\ref{fig:interaction-comparison}.}
\label{fig:late-time-scaling}
\end{center}
\end{figure}

\begin{figure*}
\begin{center}
\includegraphics{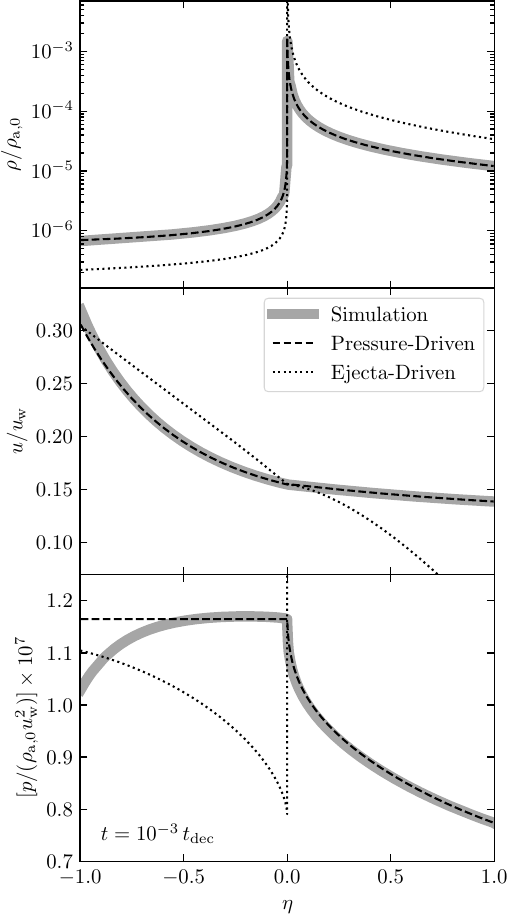}
\includegraphics{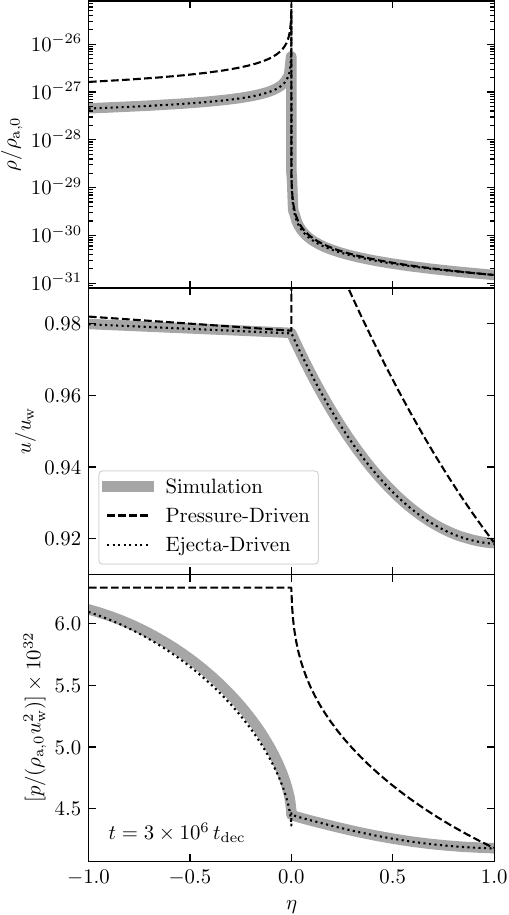}
\caption{Comparisons of our simulated fluid profiles to those of the pressure-driven and ejecta-driven solutions for $t \ll t_{\rm dec}$ (\textbf{left}) and $t \gg t_{\rm dec}$ (\textbf{right}), plotted against $\eta = (r - R_{\mathrm{c}}) / (R_{\mathrm{c}} - R_{\mathrm{r}})$ in the reverse-shocked shell and $\eta = (r - R_{\mathrm{c}}) / (R_{\mathrm{s}} - R_{\mathrm{c}})$ in the forward-shocked shell. The initial underdensity for this simulation is $f_0=10^{-3}$, and the power-law index of the ambient medium is $n=2.5$. For $t \ll t_{\rm dec}$, $f(R_{\rm c}) \ll 1$ and the pressure-driven solutions generally match the simulated profiles (with some deviation near the reverse shock due to the breakdown of isobaricity). The ejecta-driven solutions are a much better fit for $t \gg t_{\rm dec}$. Because $t_{\rm dec} \gg t_{\rm dyn}$ and the wind already starts out substantially underdense, the shocked shell densities and pressures are exceedingly low for $t \gg t_{\rm dec}$. While such small densities may not be very physically reasonable for some of the systems considered in later sections, we present these results to illustrate the agreement between the ejecta-driven solutions and the simulation output in the ejecta-driven regime.}
\label{fig:simulation-comparison}
\end{center}
\end{figure*}

The corresponding forward- and reverse-shocked shell thicknesses further clarify the origin of this transition. Defining the relative shell thicknesses as $\Delta_{\rm s} = R_{\rm s} / R_{\rm c} - 1$ and $\Delta_{\rm r} = 1 - R_{\rm r} / R_{\rm c}$, Figure~\ref{fig:late-time-scaling} shows $\Delta_{\rm s}$ and $\Delta_{\rm r}$ for the same initially accelerating simulations. Since the ambient density falls off more rapidly than the freely expanding wind density, the density contrast $f(R_{\rm c})$ increases as the bubble expands. As a result, the reverse-shocked wind region contracts relative to the overall shock structure, and the system gradually approaches the ejecta-driven limit. Deep in this limit, $\Delta_{\rm r}$ settles into the scaling $\Delta_{\rm r} \sim [f(R_{\rm c})]^{-1/2}$. At the same time, the forward-shocked shell thickness $\Delta_{\rm s}$ approaches the constant value predicted by the coasting solutions. To further demonstrate this transition, Figure~\ref{fig:simulation-comparison} compares the numerical and analytical profiles at $t \ll t_{\rm dec}$ and at $t \gg t_{\rm dec}$ as functions of $\eta$, where $\eta = (r - R_{\mathrm{c}}) / (R_{\mathrm{c}} - R_{\mathrm{r}})$ in the reverse-shocked shell, and $\eta = (r - R_{\mathrm{c}}) / (R_{\mathrm{s}} - R_{\mathrm{c}})$ in the forward-shocked shell.

\begin{figure}
\begin{center}
\includegraphics{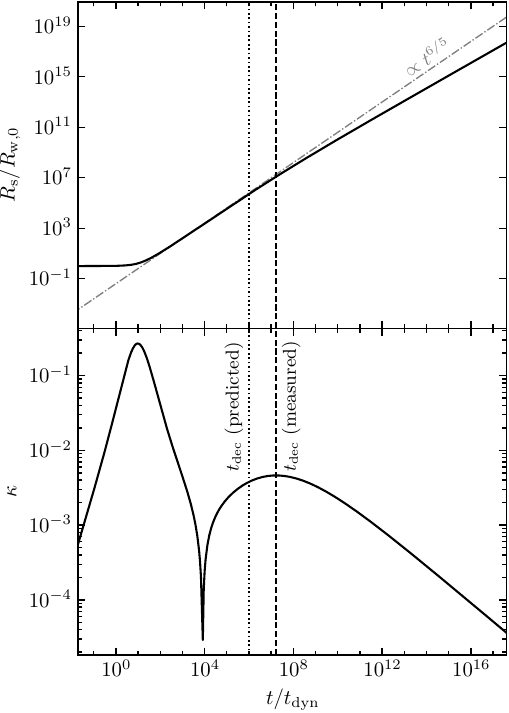}
\caption{Curvature time series of the forward shock position in log-log space for $n=2.5$ and $f_0=10^{-3}$. The pressure-driven prediction yields $R_{\rm s}/R_{\rm w,0} \propto t^{6/5}$, shown in the upper panel as a semi-transparent dash-dotted line. There is a global maximum at $t \sim$ several dynamical times corresponding to the transition from the initial transient state to the accelerating phase, followed by a secondary relative maximum associated with the onset of the coasting regime. We take the time at which this secondary maximum occurs to be the simulation $t_{\mathrm{dec}}$, shown here as the vertical dashed line. The predicted value of $t_{\rm dec}$ from Equation \ref{eq:deceleration-time} is shown as a vertical dotted line.}
\label{fig:curvature-measurement}
\end{center}
\end{figure}

We measure $t_{\mathrm{dec}}$ from our simulations by finding the point of relative maximal curvature in log-log space of the forward shock position time series. Because the curvature is technically first maximized when the system transitions from its initial transient state to the accelerating phase, we use the secondary maximum to identify $t_{\rm dec}$. An example of this measurement is shown in Figure~\ref{fig:curvature-measurement} for $f_0=10^{-3}$ and $n=2.5$.

\begin{figure}
\begin{center}
\includegraphics{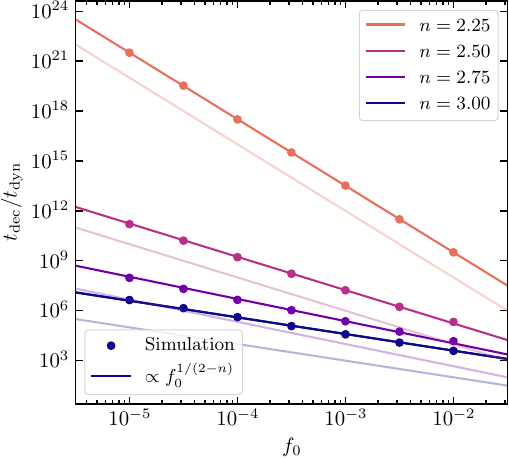}
\caption{Comparison of our measured $t_{\rm dec}$ values to the predictions from Equation~\ref{eq:deceleration-time}. The partially transparent lines represent $f_0^{1/(2-n)}$ exactly, while the opaque lines represent this same power-law but rescaled by a constant factor to more clearly demonstrate the equivalent scaling with $f_0$ between this prediction and our measurements. The values by which each curve is rescaled are different for each $n$, corresponding to $[32.27, 17.28, 23.68, 39.93]$ for $n=[2.25, 2.50, 2.75, 3.00]$.}
\label{fig:timescales}
\end{center}
\end{figure}

Figure~\ref{fig:timescales} shows $t_{\rm dec}$ as a function of $f_0$ for each value of $n$ considered, and in each case the simulations follow the predicted dependence $t_{\rm dec}\propto f_0^{1/(2-n)}$. The partially transparent curves show this scaling directly, while the opaque curves have been rescaled by constant factors to highlight the agreement in slope. The need for an order-unity normalization offset is not surprising, since Equation~\eqref{eq:deceleration-time} was obtained by setting $f(R_{\rm c})\sim 1$ and using only the leading pressure-driven scaling for the shock radius. Nevertheless, the agreement in the power-law dependence confirms the basic interpretation that the duration of the accelerating, pressure-driven stage is controlled by the time required for the increasing wind-to-ambient density contrast to reach order-unity values. The dependence on $n$ is especially strong near $n=2$, where the exponent $1/(2-n)$ becomes large. Consequently, even modest changes in $f_0$ can produce very large changes in the duration of the accelerating phase for density profiles only slightly steeper than $r^{-2}$.

\begin{figure}
\begin{center}
\includegraphics{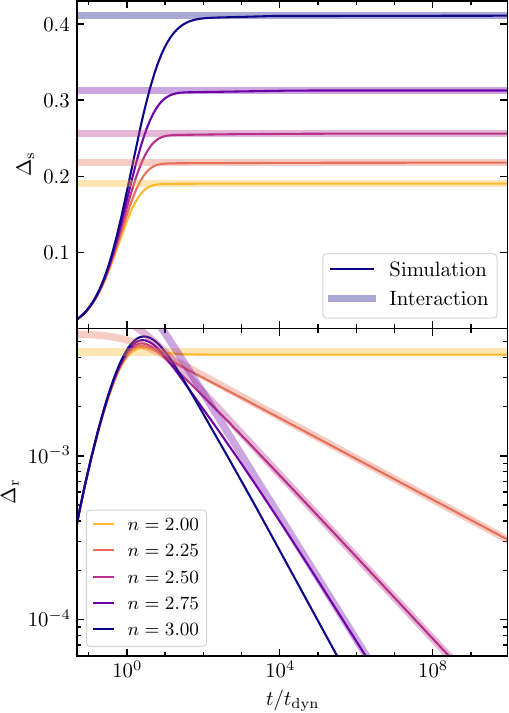}
\caption{The relative thicknesses of the forward- and reverse-shocked shells ($\Delta_{\rm s}$ and $\Delta_{\rm r}$, respectively) vs. time for $f_0 = 10^{3}$. These simulations all quickly relax directly into the coasting regime. Note that the solutions presented in \cite{2024ApJ...975L..14C} do not exist for $n=3$.}
\label{fig:interaction-comparison}
\end{center}
\end{figure}

We finally compare these results to simulations initialized with $f_0=10^3$ in Figure~\ref{fig:interaction-comparison}. Unlike the initially accelerating simulations, these systems never pass through an extended accelerating phase since they start out ejecta-driven. The reverse-shocked wind is confined to a narrow layer throughout the duration of the system's evolution, and thus $\Delta_{\rm r}$ and $\Delta_{\rm s}$ directly and robustly approach the interaction solutions. This behavior is consistent with the arguments in Section~\ref{sec:basic-considerations}: for $n>2$, once a bubble is ejecta-driven, continued expansion makes $f(R_{\rm c})$ grow with time and $\delta$ smaller with time, so the coasting approximation becomes increasingly accurate.

\section{Discussion}
\label{sec:discussion}

\subsection{Superbubble Breakout}
\label{subsec:superbubble-breakout}

Clustered stellar winds and supernovae can inflate hot bubbles in the galactic disk that `break out' at heights comparable to the disk scale height. The leading portions of the shocked shell encounter the rapidly declining density of the stratified atmosphere and accelerate away from the midplane, promoting the Rayleigh-Taylor instability and the subsequent venting of hot gas into the halo \citep{1988ApJ...324..776M, 1989ApJ...337..141M, 1992ApJ...388..103K, 2013A&A...557A.140B}. Qualitatively, these superbubble breakout events are relevant to the accelerating shock scenarios explored in this work.

Simulations indicate that this acceleration does occur, but that pre-breakout evolution is substantially affected by radiative losses and mixing \citep{2013MNRAS.434.3572R,2018MNRAS.481.3325F}. The radius at which the forward-shocked shell is expected to be cooling-dominated is \citep{Amend_Lagomarsino_Coughlin_Zrake_2026}
\begin{equation}
    \frac{R_{\mathrm{s}}}{R_{w,0}} \sim \left[ \frac{27(\gamma-1)^3L_w}{4\pi n_{a,0}^2R_{w,0}^3(1 - \xi_{\mathrm{c}}^3)\Lambda(T_{\mathrm{s}})(9\gamma-n-4)(\gamma+1)^2} \right]^{1/(3-2n)}\,,
    \label{eq:radiative-transition}
\end{equation}
where $n_{\rm a,0}$ is the ambient medium density normalization, $\Lambda$ is the temperature-dependent cooling function, and $T_{\rm s}$ is the temperature in the post-forward-shocked material. Taking $n=0$, $n_{\rm a,0}=1\,\mathrm{cm}^{-3}$, $\gamma=5/3$, $R_{\rm w,0}=10^{14}\,\mathrm{cm}$, $\Lambda(T_{\rm s})=8\times10^{-22}$, and $L_{\rm w}=10^{36}\,\mathrm{erg}\,\mathrm{s}^{-1}$ yields $R_{\rm s} \sim 1\,\mathrm{pc}$, which is indeed much smaller than typical galactic disk scale heights ($\sim$ hundreds of $\mathrm{pc}$).

Furthermore, our simulations are one-dimensional with spherical symmetry, and a robust treatment of this problem would require at least two dimensions to capture the non-negligible variation in the ambient density profile on galactic latitude. As such, in spite of the qualitative overlap, a detailed analysis of superbubble breakout events is beyond the scope of this work.

\subsection{Tidal Disruption Events}
\label{subsec:tidal-disruption-events}

Radio observations of TDEs provide a direct probe of shocks driven by fast ejecta into the circumnuclear medium \citep{2016ApJ...819L..25A, 2017ApJ...837..153A, 2021NatAs...5..491H, 2023MNRAS.522.5084G, 2023MNRAS.518..847G, 2024ApJ...971..185C, 2026ApJ..1000..139A}, and many imply steep ($n>2$) external density profiles. In several such events, modeling of the synchrotron-emitting region implies non-relativistic or mildly relativistic outflows that expand at nearly constant velocity over the epochs probed by the radio data. For ASASSN-14li, \cite{2016ApJ...819L..25A} inferred a non-relativistic outflow with $v_{\rm ej} \approx 1.2\times10^{4}\,\mathrm{km}\,\mathrm{s}^{-1}$ for spherical geometry, while also identifying a steep circumnuclear density profile, $\rho \propto r^{-2.5}$, on scales $\sim 0.01\,\mathrm{pc}$. Similarly, \cite{2021ApJ...919..127C} found that the radio-emitting outflow in AT2019dsg was in roughly free expansion ($R \propto t^{0.9}$) with $v_{\rm ej} \approx 0.07c$, even though the density profile later steepened from $\rho \propto r^{-1.6}$ to $\rho \propto r^{-3.9}$. Radio observations of additional events such as AT2020opy and AT2020vwl are also broadly consistent with this same constant or near-constant velocity expansion \citep[e.g.][]{2023MNRAS.522.5084G,2023MNRAS.518..847G}, and such late-time radio detections are typically interpreted as mildly relativistic or non-relativistic outflows \citep{2020SSRv..216...81A, 2024ApJ...971..185C}.

Within the pressure-driven wind picture, these steep density profiles would imply accelerating shocks. As discussed in prior sections, for a pressure-driven bubble expanding into $\rho \propto r^{-n}$, $R_{\rm s} \propto t^{3/(5-n)}$, so $n>2$ corresponds to $\alpha = d\ln R_{\rm s}/d\ln t > 1$. The same acceleration would also make the swept-up shell susceptible to Rayleigh-Taylor growth, since a dense shocked ambient shell is being accelerated by the lighter shocked wind interior \citep{1992ApJ...388..103K}. Thus, if the TDE radio outflows were on the accelerating energy-conserving branch, one would expect both increasing shock velocities and an acceleration-driven RT-unstable shell.

The observed near-coasting expansion instead points to the ejecta-driven regime described in \cite{2024ApJ...975L..14C}, where the forward shock approximately tracks the freely-expanding ejecta as $R_{\rm s} \sim t$. This is also consistent with high expected initial overdensities $f \sim 10^5$ \citep{Amend_Lagomarsino_Coughlin_Zrake_2026} for such TDE outflows. In this limit, for $n > 2$, the coasting solutions become increasingly robust as the instantaneous overdensity increases with expansion.

\subsection{Luminous Fast Blue Optical Transients}
\label{subsec:luminous-fast-blue-optical-transients}

LFBOTs provide another possible context in which a transition from a pressure-driven to ejecta-driven state may be relevant. Radio and millimeter modeling of several events has inferred fast shocks expanding into dense media with steep density profiles, in some cases with $n \gtrsim 2.5-3$ \citep{2020ApJ...895L..23C, 2022ApJ...926..112B, 2022ApJ...932..116H, 2025ApJ...993L...6N, 2026MNRAS.549ag678P}. Within the pressure-driven interpretation, such profiles naturally correspond to accelerating forward shocks, and the inferred shock evolution of some LFBOTs does indeed suggest $\alpha > 1$. Moreover, in a few cases the inferred behavior shows an initially accelerating expansion followed by evolution closer to constant velocity. At surface-level, this resembles the qualitative pattern found in the low-$f_0$ simulations: an initially pressure-driven bubble can accelerate while propagating through a steep density gradient, but later relax toward ejecta-driven, near-coasting expansion.

The strongest accelerators are difficult to accommodate in this framework without substantial modification. For AT2023vth \citep{2026arXiv260118926S}, the inferred shock radii imply an early expansion index $\alpha \simeq 1.7$, followed by approximately constant-velocity evolution. Fiducially, this early phase would require an effective density index $n > 3$, beyond the regime in which the pressure-driven solutions can be applied. A time-dependent wind luminosity can partially relax this constraint; if $L_{\rm w} \propto t^{\eta_{\rm w} - 1}$, then a rising luminosity where $\eta_{\rm w} = 2$ allows for $\alpha \leq 2$ with $n \leq 3$, for instance. A non-single-power-law ambient medium could also have a similar effect; a shock propagating through a locally very steep density gradient, or through the outer part of a smoothly broken power-law profile, could temporarily mimic a large effective $n$ without requiring the entire circumstellar environment to follow such a steep profile over all radii, and this was recently explored (though in the context of TDE outflows) with three-dimensional simulations in \cite{2026arXiv260605944L}.

Such a density structure could also help with the apparent sharpness of the acceleration-to-coasting transition. In the single-power-law calculations presented here, the transition occurs because $f(R_c)$ changes gradually with radius, so the relaxation from the accelerating branch to the coasting branch spans many orders of magnitude in time - inconsistent with the relative abruptness of the transition seen in radio modeling. If instead the shock crosses a density break, the effective density slope and the local wind-to-ambient density contrast can change over a much narrower radial interval. An initially steep segment could drive accelerated expansion, while a subsequent shallower segment, density enhancement, or transition into a region where the wind becomes effectively overdense could move the system toward coasting more rapidly than in the idealized models.

\begin{figure}
\begin{center}
\includegraphics{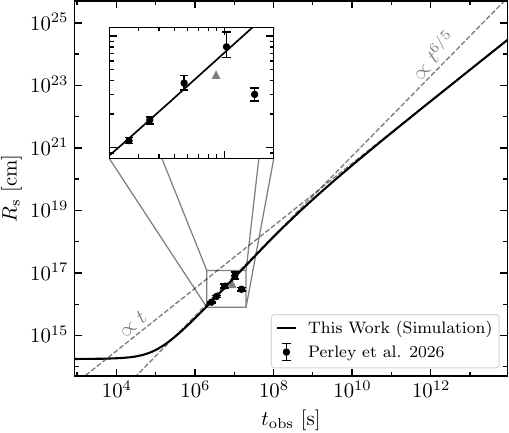}
\caption{Comparison of dimensionalised simulation output to inferred shock radius measurements of LFBOT AT2024wpp \citep{2026MNRAS.549ag678P}. This simulation is distinct from those discussed in earlier sections, and has $n=3$, $f_0=10^{-4}$, $\beta_{\rm w}=0.2$, and $R_{\rm w,0}=1.8\times 10^{14}\,\rm{cm}$. There is broad agreement with the data, consistent with mild acceleration, up until the final epoch where the shock appears to move to smaller radii.}
\label{fig:wpp}
\end{center}
\end{figure}

More mildly accelerating events are better suited to the simplified framework analysed here. AT2024wpp, for example, has been modeled as a fast, engine-driven outflow expanding into a dense environment with a steep radial density profile \citep{2025ApJ...993L...6N, 2026MNRAS.549ag678P}. Figure \ref{fig:wpp} shows the inferred shock evolution from the data presented in \citet{2026MNRAS.549ag678P}, highlighting that the early stages are consistent with mild acceleration less extreme than that of AT2023vth. The expansion index of order $\alpha \simeq 1.3-1.4$ is broadly compatible with energy-conserving evolution in a steep--but still physically plausible--ambient medium. However, this event specifically indicates a reversal of the motion of the emitting region that still cannot be captured by the framework presented here (see the discussion in the next subsection, though).

In addition to the aforementioned concerns, the pressure-driven solutions are associated with strong underdensities rather than overdensities ($f_0 \ll 1$ in our simulations). Such density contrasts are challenging to realize for plausible combinations of wind luminosity, velocity, launch radius, and circumstellar density. Taking $\rho_{\rm w} = \dot{M}/(4\pi R_{\rm w,0}^2v_{\rm w})$, then $f_0 = \dot{M}/(4\pi R_{\rm w,0}^2v_{\rm w} \rho_0)$. Adopting fiducial values of $\dot{M}=0.1\,\mathrm{M}_{\odot}\,\mathrm{yr}^{-1}$, $R_{\rm w,0} = 10^{14}\,\mathrm{cm}$, $v_{\rm w} = 0.1c$, and $\rho_0=10^{-20}\,\mathrm{g}\,\mathrm{cm}^{-3}$ yields a density contrast $\sim 10^6$. This is extremely interaction-dominated from the start, implying an initial coasting phase followed by a deceleration phase (if $n<2$, otherwise it will remain coasting), rather than an initial acceleration phase followed by a coasting phase.

A complementary possibility is that some LFBOTs are ejecta-driven throughout their observed evolution. Events like CSS161010 appear consistent with a mildly relativistic outflow, the inferred velocity of which remains roughly constant over the radio-observed interval \citep{2020ApJ...895L..23C} - a natural outcome if the system is already in the interaction regime. For $n < 2$, the system should eventually move toward the pressure-driven regime, so the lack of an observed transition could be used to place a lower limit on the transition time $t_{\rm dec}$, and hence to constrain the permitted combinations of $f_0$ and $n$, though this mapping would be highly model-dependent. It would furthermore only hold in the limit where $n < 2$; if $n > 2$, the coasting solutions are still consistent with the observed shock evolution, but the relevant parameters are degenerate 
and no such constraints can be obtained.

\subsection{Caveats in Radio Inferred Shock Dynamics}
\label{subsec:caveats}

The shock radii discussed in Sections~\ref{subsec:tidal-disruption-events} \& \ref{subsec:luminous-fast-blue-optical-transients} are inferred from synchrotron self-absorbed radio spectra using the formalism of \cite{1998ApJ...499..810C}, which is commonly applied to radio emission from TDEs and LFBOTs \citep{2016ApJ...819L..25A, 2021ApJ...919..127C, 2024ApJ...971..185C, 2023MNRAS.518..847G, 2023MNRAS.522.5084G}. The method effectively averages the properties of the emitting electrons over the width of the shocked shell, assuming an electron distribution $dN/dE \propto E^{-p}$. This is a good approximation when the emission is dominated by the forward-shocked region, where the fluid properties vary only weakly with position, and when the shell is narrow relative to the forward-shock radius, as is the case at early times in the self-similar solutions of \citep{1982ApJ...258..790C}. The inferred radius therefore characterizes the radio-emitting shell rather than directly measuring the forward-shock position, and its dependence on the fitted and assumed parameters is
\begin{equation}
    R_{\rm s} \propto \left[ \left( \frac{\epsilon_{\rm e}}{\epsilon_{\rm B}} \right)^{-1}f_V^{-1}F_{\rm p}^{p+6}(1+z)^{-(p+6)}D^{2p+12} \right]^{1/(2p+13)} \nu_{\rm p}^{-1}\,,
\end{equation}
where $F_{\rm p}$ and $\nu_{\rm p}$ are the fitted peak flux density and frequency, respectively, $\epsilon_{\rm e}$ and $\epsilon_{\rm B}$ are the electron and magnetic-field energy fractions, $D$ is the luminosity distance to the source, and $f_{V}$ is the emitting-volume filling factor. Fixed choices of microphysical parameters or geometry primarily change the normalization of $R_{\rm s}$ and do not alter the inferred expansion index $\alpha=d\ln{R_{\rm s}}/d\ln{t}$ when applied consistently across epochs. The inferred dynamics are instead most sensitive to epoch-dependent errors in $F_{\rm p}$ and $\nu_{\rm p}$, or to evolution in the assumed microphysics or emitting geometry (though perhaps less sensitively).

Analyses differ in whether the spectral slopes and turnover shape are fixed or fitted independently at each epoch \citep{2020ApJ...895L..23C, 2026MNRAS.549ag678P, 2026arXiv260118926S}. These choices can shift $F_{\rm p}$ and $\nu_{\rm p}$, particularly when the turnover is sparsely sampled, while non-simultaneous observations can distort the reconstructed spectrum if the source evolves between measurements. Such effects can bias the inferred value of the expansion index $\alpha$ and the apparent timescale over which the expansion changes.

The hydrodynamic and emitting structures of TDEs and LFBOTs may also be more complex than the homogeneous, single-zone source assumed in these fits. Multiple shocked regions or asphericity, for example, could cause the dominant radio-emitting region to change with time, in which case successive fitted radii would not necessarily trace the same physical surface. This may be relevant to the final epoch of AT2024wpp, where the final inferred radius is smaller than that of the previous epoch, which seems difficult to realize physically (i.e., while the reverse shock or contact could conceivably move to smaller radii in the rest frame of the ambient medium, how the forward shock could do so is unclear). Changes in the emitting region or spectral model could likewise affect the apparent sharpness of the acceleration-to-coasting transition inferred for AT2023vth.

The comparisons in Sections~\ref{subsec:tidal-disruption-events} \& \ref{subsec:luminous-fast-blue-optical-transients} should therefore be interpreted as qualitative tests of accelerating versus coasting evolution. Well-sampled, consistently modeled radius histories can distinguish these broad behaviors, but precise constraints on $\alpha$, $n$, $f_0$, or $t_{\rm dec}$ remain model-dependent. A more robust source-specific comparison could be achieved by jointly modeling the hydrodynamics and radio emission -- e.g., by forward-modeling radio spectra by post-processing hydrodynamic models -- rather than directly identifying the fitted emitting radius with the forward shock.

\section{Conclusions}
\label{sec:conclusions}

We have studied the adiabatic evolution of wind-driven bubbles expanding through steep ($2 \leq n \leq 3$) ambient density profiles. In the pressure-driven limit, this yields $R_{\rm s} \propto t^{3/(5-n)}$, corresponding to constant-velocity expansion for $n=2$ and acceleration for $n>2$; in contrast, the ejecta-driven solutions evolve toward constant-velocity expansion in all environments. We used one-dimensional shock-capturing hydrodynamic simulations to determine which behavior is realized for systems initialized either in the pressure-driven or ejecta-driven regime. Our main conclusions are:

\begin{enumerate}
    \item Winds initialized in the pressure-driven regime ($f_0 \ll 1$) produce initially accelerating shocks with expansion indices $\alpha > 1$, where $R_{\rm s} \propto t^{\alpha}$. As the shock expands, the density contrast $f(R_{\rm c}) = \rho_{\rm w}(R_{\rm c}) / \rho_{\rm a}(R_{\rm c})$ increases and the system evolves toward the ejecta-driven state. The transition to $\sim$ coasting expansion occurs when $f(R_{\rm c}) \sim $ a few, with $t_{\rm dec} \sim t_{\rm dyn} f_0^{1/(2-n)}$ for $n > 2$. We validated these predictions with measurements from our hydrodynamic simulations.
    \item Winds initialized in the ejecta-driven regime ($f_0 \gg 1$) do not develop an extended accelerating stage in these steep density profiles. For $n=2$, $f(R_{\rm c})$ remains constant, while for $n>2$, $f(R_{\rm c})$ increases with radius, driving the system even deeper into the coasting regime. The reverse-shocked wind shell remains thin, with $\Delta_{\rm r} \sim f^{-1/2}$, and the forward-shocked shell approaches the constant relative thickness predicted by the interaction solutions.
    \item Nearly coasting TDE radio outflows in steep circumnuclear density profiles are naturally described by the ejecta-driven solutions. In this interpretation, the roughly constant-velocity expansion of the emitting region is consistent with a highly overdense outflow impacting a steep ambient medium with $2 \leq n < 3$.
    \item LFBOTs exhibit a much broader range of behaviors. Mildly accelerating shocks followed by $\sim$ coasting evolution may be compatible with an outflow that begins in the pressure-driven regime and later approaches the ejecta-driven regime. Stronger accelerators -- i.e., events requiring expansion indices $\alpha \gtrsim 1.5$ -- are harder to describe with constant-luminosity winds in a single power-law medium with $n \leq 3$. The sharpness of some inferred acceleration-to-coasting transitions is also difficult to reproduce in this simplified setup, since the simulated transition window spans many orders of magnitude in time. Furthermore, in order for the winds to start out pressure-driven, it must be the case that $f_0 \ll 1$, which requires extreme parameters that may not be physically reasonable. Such events then likely require additional ingredients, such as time-dependent energy injection, a broken or otherwise non-power-law ambient density profile, or a different outflow geometry.
    \item Modeling uncertainties are ubiquitous in radio observations of TDEs and LFBOTs, making robust shock parameter inferences difficult. An alternative, and perhaps more self-consistent approach, could be to forward-model radio spectra from hydrodynamic simulations instead.
\end{enumerate}

The calculations presented in this work assume a constant-luminosity, spherical wind expanding through a single power-law ambient medium, and relaxing these assumptions would be a natural extension of this work. Such extensions would be particularly relevant for source-specific applications to LFBOTs and some TDE outflows.

\section*{Acknowledgements}
B.A. and E.R.C.~acknowledge support from NASA through the Astrophysics Theory Program, grant 80NSSC24K0897, and through Chandra Award Number 25700383 issued by the Chandra X-ray Observatory Center, which is operated by the Smithsonian Astrophysical Observatory for and on behalf of the National Aeronautics Space Administration under contract NAS8-03060.
M.C. and A.Y.Q.H. acknowledge support from a Sloan Research Fellowship (Award Number FG-2024-21320) from the Alfred P. Sloan Foundation, and a Packard Fellowship from the David and Lucile Packard Foundation.

\bibliography{references}{}

@ARTICLE{1972SvA....15..708A,
       author = {{Avedisova}, V.~S.},
        title = "{Formation of Nebulae by Wolf-Rayet Stars.}",
      journal = {\sovast},
         year = 1972,
        month = apr,
       volume = {15},
        pages = {708},
       adsurl = {https://ui.adsabs.harvard.edu/abs/1972SvA....15..708A}
}

@ARTICLE{2016ApJ...819L..25A,
       author = {{Alexander}, K.~D. and {Berger}, E. and {Guillochon}, J. and {Zauderer}, B.~A. and {Williams}, P.~K.~G.},
        title = "{Discovery of an Outflow from Radio Observations of the Tidal Disruption Event ASASSN-14li}",
      journal = {\apjl},
         year = 2016,
        month = mar,
       volume = {819},
       number = {2},
          eid = {L25},
        pages = {L25},
          doi = {10.3847/2041-8205/819/2/L25},
archivePrefix = {arXiv},
       eprint = {1510.01226},
 primaryClass = {astro-ph.HE},
       adsurl = {https://ui.adsabs.harvard.edu/abs/2016ApJ...819L..25A}
}

@ARTICLE{2020SSRv..216...81A,
       author = {{Alexander}, Kate D. and {van Velzen}, Sjoert and {Horesh}, Assaf and {Zauderer}, B. Ashley},
        title = "{Radio Properties of Tidal Disruption Events}",
      journal = {\ssr},
         year = 2020,
        month = jun,
       volume = {216},
       number = {5},
          eid = {81},
        pages = {81},
          doi = {10.1007/s11214-020-00702-w},
archivePrefix = {arXiv},
       eprint = {2006.01159},
 primaryClass = {astro-ph.HE},
       adsurl = {https://ui.adsabs.harvard.edu/abs/2020SSRv..216...81A}
}

@article{Amend_Lagomarsino_Coughlin_Zrake_2026, title={The interaction phase of engine-driven explosions and high-energy winds}, DOI={10.1017/pasa.2026.10235}, journal={Publications of the Astronomical Society of Australia}, author={Amend, Benjamin and Lagomarsino, Christopher and Coughlin, Eric R. and Zrake, Jonathan}, year={2026}, pages={1–17}}

@ARTICLE{2017ApJ...837..153A,
       author = {{Alexander}, K.~D. and {Wieringa}, M.~H. and {Berger}, E. and {Saxton}, R.~D. and {Komossa}, S.},
        title = "{Radio Observations of the Tidal Disruption Event XMMSL1 J0740-85}",
      journal = {\apj},
         year = 2017,
        month = mar,
       volume = {837},
       number = {2},
          eid = {153},
        pages = {153},
          doi = {10.3847/1538-4357/aa6192},
archivePrefix = {arXiv},
       eprint = {1610.03861},
 primaryClass = {astro-ph.HE},
       adsurl = {https://ui.adsabs.harvard.edu/abs/2017ApJ...837..153A}
}

@ARTICLE{2026ApJ..1000..139A,
       author = {{Alexander}, Kate D. and {Margutti}, Raffaella and {Gomez}, Sebastian and {Stroh}, Michael and {Chornock}, Ryan and {Laskar}, Tanmoy and {Cendes}, Y. and {Berger}, Edo and {Eftekhari}, Tarraneh and {Franz}, Noah and et al.},
        title = "{The Multiwavelength Context of Delayed Radio Emission in Tidal Disruption Events: Evidence for Accretion-driven Outflows}",
      journal = {\apj},
         year = 2026,
        month = mar,
       volume = {1000},
       number = {1},
          eid = {139},
        pages = {139},
          doi = {10.3847/1538-4357/ae40ab},
archivePrefix = {arXiv},
       eprint = {2506.12729},
 primaryClass = {astro-ph.HE},
       adsurl = {https://ui.adsabs.harvard.edu/abs/2026ApJ..1000..139A}
}

@ARTICLE{2022ApJ...926..112B,
       author = {{Bright}, Joe S. and {Margutti}, Raffaella and {Matthews}, David and {Brethauer}, Daniel and {Coppejans}, Deanne and {Wieringa}, Mark H. and {Metzger}, Brian D. and {DeMarchi}, Lindsay and {Laskar}, Tanmoy and {Romero}, Charles and {Alexander}, Kate D. and {Horesh}, Assaf and {Migliori}, Giulia and {Chornock}, Ryan and {Berger}, E. and {Bietenholz}, Michael and {Devlin}, Mark J. and {Dicker}, Simon R. and {Jacobson-Gal{\'a}n}, W.~V. and {Mason}, Brian S. and {Milisavljevic}, Dan and {Motta}, Sara E. and {Mroczkowski}, Tony and {Ramirez-Ruiz}, Enrico and {Rhodes}, Lauren and {Sarazin}, Craig L. and {Sfaradi}, Itai and {Sievers}, Jonathan},
        title = "{Radio and X-Ray Observations of the Luminous Fast Blue Optical Transient AT 2020xnd}",
      journal = {\apj},
         year = 2022,
        month = feb,
       volume = {926},
       number = {2},
          eid = {112},
        pages = {112},
          doi = {10.3847/1538-4357/ac4506},
archivePrefix = {arXiv},
       eprint = {2110.05514},
 primaryClass = {astro-ph.HE},
       adsurl = {https://ui.adsabs.harvard.edu/abs/2022ApJ...926..112B}
}

@ARTICLE{2013A&A...557A.140B,
       author = {{Baumgartner}, V. and {Breitschwerdt}, D.},
        title = "{Superbubble evolution in disk galaxies. I. Study of blow-out by analytical models}",
      journal = {\aap},
         year = 2013,
        month = sep,
       volume = {557},
          eid = {A140},
        pages = {A140},
          doi = {10.1051/0004-6361/201321261},
archivePrefix = {arXiv},
       eprint = {1402.0194},
 primaryClass = {astro-ph.GA},
       adsurl = {https://ui.adsabs.harvard.edu/abs/2013A&A...557A.140B}
}

@ARTICLE{1975ApJ...195..157C,
       author = {{Castor}, J.~I. and {Abbott}, D.~C. and {Klein}, R.~I.},
        title = "{Radiation-driven winds in Of stars.}",
      journal = {\apj},
         year = 1975,
        month = jan,
       volume = {195},
        pages = {157-174},
          doi = {10.1086/153315},
       adsurl = {https://ui.adsabs.harvard.edu/abs/1975ApJ...195..157C}
}

@ARTICLE{2007ARA&A..45..177C,
       author = {{Crowther}, Paul A.},
        title = "{Physical Properties of Wolf-Rayet Stars}",
      journal = {\araa},
         year = 2007,
        month = sep,
       volume = {45},
       number = {1},
        pages = {177-219},
          doi = {10.1146/annurev.astro.45.051806.110615},
archivePrefix = {arXiv},
       eprint = {astro-ph/0610356},
 primaryClass = {astro-ph},
       adsurl = {https://ui.adsabs.harvard.edu/abs/2007ARA&A..45..177C}
}

@ARTICLE{1985Natur.317...44C,
       author = {{Chevalier}, R.~A. and {Clegg}, A.~W.},
        title = "{Wind from a starburst galaxy nucleus}",
      journal = {\nat},
         year = 1985,
        month = sep,
       volume = {317},
       number = {6032},
        pages = {44-45},
          doi = {10.1038/317044a0},
       adsurl = {https://ui.adsabs.harvard.edu/abs/1985Natur.317...44C}
}

@ARTICLE{1975ApJ...200L.107C,
       author = {{Castor}, J. and {McCray}, R. and {Weaver}, R.},
        title = "{Interstellar bubbles.}",
      journal = {\apjl},
         year = 1975,
        month = sep,
       volume = {200},
        pages = {L107-L110},
          doi = {10.1086/181908},
       adsurl = {https://ui.adsabs.harvard.edu/abs/1975ApJ...200L.107C}
}

@ARTICLE{2024ApJ...975L..14C,
       author = {{Coughlin}, Eric R.},
        title = "{From Coasting to Energy-conserving: New Self-similar Solutions to the Interaction Phase of Strong Explosions}",
      journal = {\apjl},
         year = 2024,
        month = nov,
       volume = {975},
       number = {1},
          eid = {L14},
        pages = {L14},
          doi = {10.3847/2041-8213/ad87cc},
archivePrefix = {arXiv},
       eprint = {2409.10600},
 primaryClass = {astro-ph.HE},
       adsurl = {https://ui.adsabs.harvard.edu/abs/2024ApJ...975L..14C}
}

@ARTICLE{2024ApJ...971..185C,
       author = {{Cendes}, Y. and {Berger}, E. and {Alexander}, K.~D. and {Chornock}, R. and {Margutti}, R. and {Metzger}, B. and {Wieringa}, M.~H. and {Bietenholz}, M.~F. and {Hajela}, A. and {Laskar}, T. and et al.},
        title = "{Ubiquitous Late Radio Emission from Tidal Disruption Events}",
      journal = {\apj},
         year = 2024,
        month = aug,
       volume = {971},
       number = {2},
          eid = {185},
        pages = {185},
          doi = {10.3847/1538-4357/ad5541},
archivePrefix = {arXiv},
       eprint = {2308.13595},
 primaryClass = {astro-ph.HE},
       adsurl = {https://ui.adsabs.harvard.edu/abs/2024ApJ...971..185C}
}

@ARTICLE{2020ApJ...895L..23C,
       author = {{Coppejans}, D.~L. and {Margutti}, R. and {Terreran}, G. and {Nayana}, A.~J. and {Coughlin}, E.~R. and {Laskar}, T. and {Alexander}, K.~D. and {Bietenholz}, M. and {Caprioli}, D. and {Chandra}, P. and et al.},
        title = "{A Mildly Relativistic Outflow from the Energetic, Fast-rising Blue Optical Transient CSS161010 in a Dwarf Galaxy}",
      journal = {\apjl},
         year = 2020,
        month = may,
       volume = {895},
       number = {1},
          eid = {L23},
        pages = {L23},
          doi = {10.3847/2041-8213/ab8cc7},
archivePrefix = {arXiv},
       eprint = {2003.10503},
 primaryClass = {astro-ph.HE},
       adsurl = {https://ui.adsabs.harvard.edu/abs/2020ApJ...895L..23C}
}

@ARTICLE{2021ApJ...919..127C,
       author = {{Cendes}, Y. and {Alexander}, K.~D. and {Berger}, E. and {Eftekhari}, T. and {Williams}, P.~K.~G. and {Chornock}, R.},
        title = "{Radio Observations of an Ordinary Outflow from the Tidal Disruption Event AT2019dsg}",
      journal = {\apj},
         year = 2021,
        month = oct,
       volume = {919},
       number = {2},
          eid = {127},
        pages = {127},
          doi = {10.3847/1538-4357/ac110a},
archivePrefix = {arXiv},
       eprint = {2103.06299},
 primaryClass = {astro-ph.HE},
       adsurl = {https://ui.adsabs.harvard.edu/abs/2021ApJ...919..127C}
}

@ARTICLE{1982ApJ...258..790C,
       author = {{Chevalier}, R.~A.},
        title = "{Self-similar solutions for the interaction of stellar ejecta with an external medium.}",
      journal = {\apj},
         year = 1982,
        month = jul,
       volume = {258},
        pages = {790-797},
          doi = {10.1086/160126},
       adsurl = {https://ui.adsabs.harvard.edu/abs/1982ApJ...258..790C}
}

@ARTICLE{2018MNRAS.481.3325F,
       author = {{Fielding}, Drummond and {Quataert}, Eliot and {Martizzi}, Davide},
        title = "{Clustered supernovae drive powerful galactic winds after superbubble breakout}",
      journal = {\mnras},
         year = 2018,
        month = dec,
       volume = {481},
       number = {3},
        pages = {3325-3347},
          doi = {10.1093/mnras/sty2466},
archivePrefix = {arXiv},
       eprint = {1807.08758},
 primaryClass = {astro-ph.GA},
       adsurl = {https://ui.adsabs.harvard.edu/abs/2018MNRAS.481.3325F}
}

@ARTICLE{2012MNRAS.425..605F,
       author = {{Faucher-Gigu{\`e}re}, Claude-Andr{\'e} and {Quataert}, Eliot},
        title = "{The physics of galactic winds driven by active galactic nuclei}",
      journal = {\mnras},
         year = 2012,
        month = sep,
       volume = {425},
       number = {1},
        pages = {605-622},
          doi = {10.1111/j.1365-2966.2012.21512.x},
archivePrefix = {arXiv},
       eprint = {1204.2547},
 primaryClass = {astro-ph.CO},
       adsurl = {https://ui.adsabs.harvard.edu/abs/2012MNRAS.425..605F}
}

@ARTICLE{2021ARA&A..59...21G,
       author = {{Gezari}, Suvi},
        title = "{Tidal Disruption Events}",
      journal = {\araa},
         year = 2021,
        month = sep,
       volume = {59},
        pages = {21-58},
          doi = {10.1146/annurev-astro-111720-030029},
archivePrefix = {arXiv},
       eprint = {2104.14580},
 primaryClass = {astro-ph.HE},
       adsurl = {https://ui.adsabs.harvard.edu/abs/2021ARA&A..59...21G}
}

@ARTICLE{2023MNRAS.522.5084G,
       author = {{Goodwin}, A.~J. and {Alexander}, K.~D. and {Miller-Jones}, J.~C.~A. and {Bietenholz}, M.~F. and {van Velzen}, S. and {Anderson}, G.~E. and {Berger}, E. and {Cendes}, Y. and {Chornock}, R. and {Coppejans}, D.~L. and et al.},
        title = "{A radio-emitting outflow produced by the tidal disruption event AT2020vwl}",
      journal = {\mnras},
         year = 2023,
        month = jul,
       volume = {522},
       number = {4},
        pages = {5084-5097},
          doi = {10.1093/mnras/stad1258},
archivePrefix = {arXiv},
       eprint = {2304.12661},
 primaryClass = {astro-ph.HE},
       adsurl = {https://ui.adsabs.harvard.edu/abs/2023MNRAS.522.5084G}
}

@ARTICLE{2023MNRAS.518..847G,
       author = {{Goodwin}, A.~J. and {Miller-Jones}, J.~C.~A. and {van Velzen}, S. and {Bietenholz}, M. and {Greenland}, J. and {Cenko}, B. and {Gezari}, S. and {Horesh}, A. and {Sivakoff}, G.~R. and {Yan}, L. and et al.},
        title = "{Radio observations of the tidal disruption event AT2020opy: a luminous non-relativistic outflow encountering a dense circumnuclear medium}",
      journal = {\mnras},
         year = 2023,
        month = jan,
       volume = {518},
       number = {1},
        pages = {847-854},
          doi = {10.1093/mnras/stac3127},
archivePrefix = {arXiv},
       eprint = {2208.13967},
 primaryClass = {astro-ph.HE},
       adsurl = {https://ui.adsabs.harvard.edu/abs/2023MNRAS.518..847G}
}

@ARTICLE{2021NatAs...5..491H,
       author = {{Horesh}, A. and {Cenko}, S.~B. and {Arcavi}, I.},
        title = "{Delayed radio flares from a tidal disruption event}",
      journal = {Nature Astronomy},
         year = 2021,
        month = may,
       volume = {5},
        pages = {491-497},
          doi = {10.1038/s41550-021-01300-8},
archivePrefix = {arXiv},
       eprint = {2102.11290},
 primaryClass = {astro-ph.HE},
       adsurl = {https://ui.adsabs.harvard.edu/abs/2021NatAs...5..491H}
}

@ARTICLE{2022ApJ...932..116H,
       author = {{Ho}, Anna Y.~Q. and {Margalit}, Ben and {Bremer}, Michael and {Perley}, Daniel A. and {Yao}, Yuhan and {Dobie}, Dougal and {Kaplan}, David L. and {O'Brien}, Andrew and {Petitpas}, Glen and {Zic}, Andrew},
        title = "{Luminous Millimeter, Radio, and X-Ray Emission from ZTF 20acigmel (AT 2020xnd)}",
      journal = {\apj},
         year = 2022,
        month = jun,
       volume = {932},
       number = {2},
          eid = {116},
        pages = {116},
          doi = {10.3847/1538-4357/ac4e97},
archivePrefix = {arXiv},
       eprint = {2110.05490},
 primaryClass = {astro-ph.HE},
       adsurl = {https://ui.adsabs.harvard.edu/abs/2022ApJ...932..116H}
}

@ARTICLE{2015ARA&A..53..115K,
       author = {{King}, Andrew and {Pounds}, Ken},
        title = "{Powerful Outflows and Feedback from Active Galactic Nuclei}",
      journal = {\araa},
         year = 2015,
        month = aug,
       volume = {53},
        pages = {115-154},
          doi = {10.1146/annurev-astro-082214-122316},
archivePrefix = {arXiv},
       eprint = {1503.05206},
 primaryClass = {astro-ph.GA},
       adsurl = {https://ui.adsabs.harvard.edu/abs/2015ARA&A..53..115K}
}

@ARTICLE{1992ApJ...388..103K,
       author = {{Koo}, Bon-Chul and {McKee}, Christopher F.},
        title = "{Dynamics of Wind Bubbles and Superbubbles. II. Analytic Theory}",
      journal = {\apj},
         year = 1992,
        month = mar,
       volume = {388},
        pages = {103},
          doi = {10.1086/171133},
       adsurl = {https://ui.adsabs.harvard.edu/abs/1992ApJ...388..103K}
}

@ARTICLE{2017ApJ...834...25K,
       author = {{Kim}, Chang-Goo and {Ostriker}, Eve C. and {Raileanu}, Roberta},
        title = "{Superbubbles in the Multiphase ISM and the Loading of Galactic Winds}",
      journal = {\apj},
         year = 2017,
        month = jan,
       volume = {834},
       number = {1},
          eid = {25},
        pages = {25},
          doi = {10.3847/1538-4357/834/1/25},
archivePrefix = {arXiv},
       eprint = {1610.03092},
 primaryClass = {astro-ph.GA},
       adsurl = {https://ui.adsabs.harvard.edu/abs/2017ApJ...834...25K}
}

@ARTICLE{2026arXiv260605944L,
       author = {{Lei}, Xiangli and {Wu}, Qingwen and {Zhou}, Chang and {Lei}, Wei-Hua and {Li}, Ya-Ping and {Wu}, Jiancheng and {Yang}, Weibo},
        title = "{Simulations of interaction between outflow and surrounding broken power-law circumnuclear medium: implications for different radio light curves of TDEs}",
      journal = {arXiv e-prints},
         year = 2026,
        month = jun,
          eid = {arXiv:2606.05944},
        pages = {arXiv:2606.05944},
          doi = {10.48550/arXiv.2606.05944},
archivePrefix = {arXiv},
       eprint = {2606.05944},
 primaryClass = {astro-ph.HE},
       adsurl = {https://ui.adsabs.harvard.edu/abs/2026arXiv260605944L}
}

@ARTICLE{1995ApJ...451..498M,
       author = {{Murray}, N. and {Chiang}, J. and {Grossman}, S.~A. and {Voit}, G.~M.},
        title = "{Accretion Disk Winds from Active Galactic Nuclei}",
      journal = {\apj},
         year = 1995,
        month = oct,
       volume = {451},
        pages = {498},
          doi = {10.1086/176238},
       adsurl = {https://ui.adsabs.harvard.edu/abs/1995ApJ...451..498M}
}

@ARTICLE{1988ApJ...324..776M,
       author = {{Mac Low}, Mordecai-Mark and {McCray}, Richard},
        title = "{Superbubbles in Disk Galaxies}",
      journal = {\apj},
         year = 1988,
        month = jan,
       volume = {324},
        pages = {776},
          doi = {10.1086/165936},
       adsurl = {https://ui.adsabs.harvard.edu/abs/1988ApJ...324..776M}
}

@ARTICLE{1989ApJ...337..141M,
       author = {{Mac Low}, Mordecai-Mark and {McCray}, Richard and {Norman}, Michael L.},
        title = "{Superbubble Blowout Dynamics}",
      journal = {\apj},
         year = 1989,
        month = feb,
       volume = {337},
        pages = {141},
          doi = {10.1086/167094},
       adsurl = {https://ui.adsabs.harvard.edu/abs/1989ApJ...337..141M}
}

@ARTICLE{2025ApJ...993L...6N,
       author = {{Nayana}, A.~J. and {Margutti}, Raffaella and {Wiston}, Eli and {Laskar}, Tanmoy and {Migliori}, Giulia and {Chornock}, Ryan and {Galvin}, Timothy J. and {LeBaron}, Natalie and {Hajela}, Aprajita and {Christy}, Collin T. and et al.},
        title = "{The Most Luminous Known Fast Blue Optical Transient AT 2024wpp: Unprecedented Evolution and Properties in the X-Rays and Radio}",
      journal = {\apjl},
         year = 2025,
        month = nov,
       volume = {993},
       number = {1},
          eid = {L6},
        pages = {L6},
          doi = {10.3847/2041-8213/ae0b4d},
archivePrefix = {arXiv},
       eprint = {2509.00952},
 primaryClass = {astro-ph.HE},
       adsurl = {https://ui.adsabs.harvard.edu/abs/2025ApJ...993L...6N}
}

@ARTICLE{2026MNRAS.549ag678P,
       author = {{Perley}, Daniel A. and {Ho}, Anna Y.~Q. and {McGrath}, Zo{\"e} and {Camilo}, Michael and {Sevilla}, Cassie and {Chen}, Ping and {Schroeder}, Genevieve and {Govreen-Segal}, Taya and {Bochenek}, Aleksandra and {Qin}, Yu-Jing and et al.},
        title = "{AT 2024wpp: an extremely luminous fast ultraviolet transient powered by accretion onto a black hole}",
      journal = {\mnras},
         year = 2026,
        month = jun,
       volume = {549},
       number = {1},
          eid = {stag678},
        pages = {stag678},
          doi = {10.1093/mnras/stag678},
archivePrefix = {arXiv},
       eprint = {2601.03337},
 primaryClass = {astro-ph.HE},
       adsurl = {https://ui.adsabs.harvard.edu/abs/2026MNRAS.549ag678P}
}

@ARTICLE{1988Natur.333..523R,
       author = {{Rees}, Martin J.},
        title = "{Tidal disruption of stars by black holes of {}10$^{6}$-{}10$^{8}$ solar masses in nearby galaxies}",
      journal = {\nat},
         year = 1988,
        month = jun,
       volume = {333},
       number = {6173},
        pages = {523-528},
          doi = {10.1038/333523a0},
       adsurl = {https://ui.adsabs.harvard.edu/abs/1988Natur.333..523R}
}

@ARTICLE{2013MNRAS.434.3572R,
       author = {{Roy}, Arpita and {Nath}, Biman B. and {Sharma}, Prateek and {Shchekinov}, Yuri},
        title = "{Superbubble breakout and galactic winds from disc galaxies}",
      journal = {\mnras},
         year = 2013,
        month = oct,
       volume = {434},
       number = {4},
        pages = {3572-3581},
          doi = {10.1093/mnras/stt1279},
archivePrefix = {arXiv},
       eprint = {1303.2664},
 primaryClass = {astro-ph.GA},
       adsurl = {https://ui.adsabs.harvard.edu/abs/2013MNRAS.434.3572R}
}

@ARTICLE{2011MNRAS.415..168S,
       author = {{Strubbe}, Linda E. and {Quataert}, Eliot},
        title = "{Spectroscopic signatures of the tidal disruption of stars by massive black holes}",
      journal = {\mnras},
         year = 2011,
        month = jul,
       volume = {415},
       number = {1},
        pages = {168-180},
          doi = {10.1111/j.1365-2966.2011.18686.x},
archivePrefix = {arXiv},
       eprint = {1008.4131},
 primaryClass = {astro-ph.CO},
       adsurl = {https://ui.adsabs.harvard.edu/abs/2011MNRAS.415..168S}
}

@ARTICLE{2026arXiv260118926S,
       author = {{Sevilla}, Cassie and {Ho}, Anna Y.~Q. and {Nayana A.}, J. and {Schulze}, Steve and {Perley}, Daniel A. and {Bremer}, Michael and {Andreoni}, Igor and {Altunin}, Ivan and {Brink}, Thomas G. and {Chandra}, Poonam and et al.},
        title = "{Multiwavelength Analysis of Six Luminous, Fast Blue Optical Transients}",
      journal = {arXiv e-prints},
         year = 2026,
        month = jan,
          eid = {arXiv:2601.18926},
        pages = {arXiv:2601.18926},
          doi = {10.48550/arXiv.2601.18926},
archivePrefix = {arXiv},
       eprint = {2601.18926},
 primaryClass = {astro-ph.HE},
       adsurl = {https://ui.adsabs.harvard.edu/abs/2026arXiv260118926S}
}

@ARTICLE{2016MNRAS.455.1830T,
       author = {{Thompson}, Todd A. and {Quataert}, Eliot and {Zhang}, Dong and {Weinberg}, David H.},
        title = "{An origin for multiphase gas in galactic winds and haloes}",
      journal = {\mnras},
         year = 2016,
        month = jan,
       volume = {455},
       number = {2},
        pages = {1830-1844},
          doi = {10.1093/mnras/stv2428},
archivePrefix = {arXiv},
       eprint = {1507.04362},
 primaryClass = {astro-ph.GA},
       adsurl = {https://ui.adsabs.harvard.edu/abs/2016MNRAS.455.1830T}
}

@ARTICLE{2001A&A...369..574V,
       author = {{Vink}, Jorick S. and {de Koter}, A. and {Lamers}, H.~J.~G.~L.~M.},
        title = "{Mass-loss predictions for O and B stars as a function of metallicity}",
      journal = {\aap},
         year = 2001,
        month = apr,
       volume = {369},
        pages = {574-588},
          doi = {10.1051/0004-6361:20010127},
archivePrefix = {arXiv},
       eprint = {astro-ph/0101509},
 primaryClass = {astro-ph},
       adsurl = {https://ui.adsabs.harvard.edu/abs/2001A&A...369..574V}
}

@ARTICLE{2005ARA&A..43..769V,
       author = {{Veilleux}, Sylvain and {Cecil}, Gerald and {Bland-Hawthorn}, Joss},
        title = "{Galactic Winds}",
      journal = {\araa},
         year = 2005,
        month = sep,
       volume = {43},
       number = {1},
        pages = {769-826},
          doi = {10.1146/annurev.astro.43.072103.150610},
archivePrefix = {arXiv},
       eprint = {astro-ph/0504435},
 primaryClass = {astro-ph},
       adsurl = {https://ui.adsabs.harvard.edu/abs/2005ARA&A..43..769V}
}

@ARTICLE{1977ApJ...218..377W,
       author = {{Weaver}, R. and {McCray}, R. and {Castor}, J. and {Shapiro}, P. and {Moore}, R.},
        title = "{Interstellar bubbles. II. Structure and evolution.}",
      journal = {\apj},
         year = 1977,
        month = dec,
       volume = {218},
        pages = {377-395},
          doi = {10.1086/155692},
       adsurl = {https://ui.adsabs.harvard.edu/abs/1977ApJ...218..377W}
}

@ARTICLE{2005MNRAS.363...93Z,
       author = {{Zou}, Y.~C. and {Wu}, X.~F. and {Dai}, Z.~G.},
        title = "{Early afterglows in wind environments revisited}",
      journal = {\mnras},
         year = 2005,
        month = oct,
       volume = {363},
       number = {1},
        pages = {93-106},
          doi = {10.1111/j.1365-2966.2005.09411.x},
archivePrefix = {arXiv},
       eprint = {astro-ph/0508602},
 primaryClass = {astro-ph},
       adsurl = {https://ui.adsabs.harvard.edu/abs/2005MNRAS.363...93Z}
}

@ARTICLE{1998ApJ...499..810C,
       author = {{Chevalier}, Roger A.},
        title = "{Synchrotron Self-Absorption in Radio Supernovae}",
      journal = {\apj},
         year = 1998,
        month = may,
       volume = {499},
       number = {2},
        pages = {810-819},
          doi = {10.1086/305676},
       adsurl = {https://ui.adsabs.harvard.edu/abs/1998ApJ...499..810C}
}
\bibliographystyle{paslike}

\end{document}